# Tri-band Aperture-shared Antenna Array Using Scalable FSS-based Electromagnetic Transparent Structure

Yongzheng Li, *Student Member, IEEE,* Wanchen Yang, *Senior Member, IEEE,* Quan Xue, *Fellow, IEEE* and Wenquan Che, *Fellow, IEEE*

*Abstract*—In a tri-band aperture-shared array (TBA), the low-band (LB) dipole often deteriorates the radiation patterns of the middle-band (MB) and high-band (HB) antennas due to shielding effects. To address this issue, a novel dual-band electromagnetic transparent structure (DBTS) is firstly proposed and used to realize a TBA. The DBTS achieves two tunable electromagnetic transparent frequency bands by periodically loading capacitive patches and meander lines to an inductive strip. Meanwhile, the DBTS features flexible frequency band scalability by loading additional serial *L-C* tanks. Then the DBTS is utilized to construct the LB dipole of a TBA, with its electromagnetic transparent bands allocated at MB and HB simultaneously. The proposed array realizes an aperture-shared operation within the frequency ranges covering 0.65-0.88 GHz (LB), 1.92-2.18 GHz (MB) and 3.3-3.8 GHz (HB). The LB dipole induces minimal shielding to the MB and HB antennas, resulting in their restored radiation performance with broadside gain deviation of less than 0.6 dB.

*Index Terms*—Aperture-shared antenna array, base-station antenna, electromagnetic transparent structure, radiation pattern distortion.

## I. INTRODUCTION

Multi-band aperture-shared antenna arrays can save aperture space thereby reduce deployment costs in next-generation communication system. However, the LB antenna often causes strong shielding effects, degrading the radiation performance at higher bands. To address this issue, various methods have been proposed [1]–[25]. Top-bottom scheme [1-4] putting the higher band antennas over the LB antenna, with a low-pass surface inserted between them. Specifically, a TBA with wide bandwidth was proposed in [4]. However, the helix coils used in the feeding structure of the HB and MB antennas introduces additional loss. Besides, the stacked structure reduces mechanical stability. Choke loading [5]–[14] effectively suppresses induced currents on the LB radiator, but often narrows the LB bandwidth (e.g., 19.4% in [8]). Slot-ring FSS was constructed on the arm of the LB dipole in [15]–[21], but the size of LB radiator needs to be compromised with the dimension of the FSS. Many other methods, such as loading slots [22], constructing the quasi F-P cavity [23], loading cloaks [24] or dielectrics [25], can also mitigate the shielding effect, but only dual-band aperture-shared array were realized. Therefore, a tri-band aperture-shared solution is still desirable—one that maintains LB performance, supports multi-band integration, and ensures structural stability.

In this work, to minimize the shielding effect caused by the LB dipole on the MB and HB antennas in the TBA, a DBTS is proposed firstly, which is capable of producing two tunable electromagnetic transparent bands. The proposed DBTS features flexible frequency band scalability by adding *L-C* tanks to its equivalent circuit. Based on this DBTS, an LB dipole is designed and integrated with a dual-band sub-array operating at MB and HB. The resulting TBA achieves compact tri-band integration with minimal shielding, a simple structure, and scalable frequency adaptability.

## II. ELECTROMAGNETIC TRANSPARENT STRUCTURE BASED ON FSS

As illustrated in Fig. 1, the proposed DBTS is originated from one thin strip, which is equivalent to a shunt inductor $L_p$, and termed with *Case*-A. Then *Case*-B is obtained from *Case*-A by loading the periodic meander line on the opposite side of the laminate, and this introduces a serial $L_s$-$C_s$ tank in parallel with the $L_p$. *Case*-C represents the proposed DBTS, which incorporates capacitive patches on the basis of *Case*-B, these capacitive patches induce the serial tank $L_{s1}$-$C_{s1}$. Note that the capacitive patches are etched on both side of two additional laminates, the air spacing between the laminates is modeled with transmission lines. The value of $L_p$ and $C_{s1}$ are calculated from the dimensions of structure using the equations in [26], while the $L_{s1}$, $L_s$ and $C_s$ are extracted using the method in [27]. Furthermore, slight parameter tuning is performed to improve accuracy.

This work was supported in part by the National Natural Science Foundation of China under Grant 62321002. (*Corresponding author: Wenquan Che*)

Y. Li, Q. Xue and W. Che are with the Guangdong Provincial Key Laboratory of Millimeter-Wave and Terahertz, Guangdong-Hong Kong-Macao Joint Laboratory for Millimeter-Wave and Terahertz, School of Electronic and Information Engineering, South China University of Technology, Guangzhou 510641, China. (e-mail: eelyz@ieee.org, eeqxue@scut.edu.cn, eewqche@scut.edu.cn).

W. Yang is with the College of Electronic and Information Engineering, Nanjing University of Aeronautics and Astronautics, Nanjing 211106, China. (e-mail: wcyang@nuaa.edu.cn).

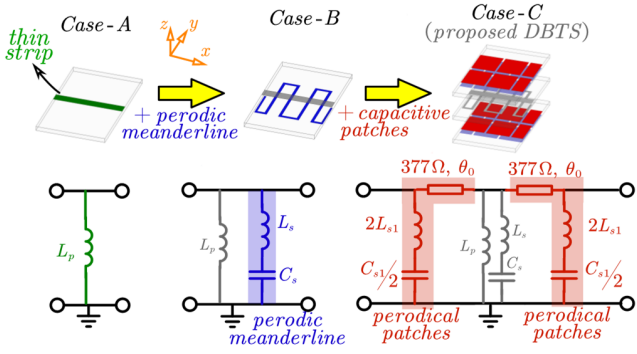

**Fig. 1.** Configuration and equivalent circuit for *Case*-A to *Case*-C. The circuit parameters: $L_{s1}$=0.1 nH, $C_{s1}$=0.28 pF, $L_p$=8.2 nH, $L_s$=45.7 nH, $C_s$=0.0915 pF, $\theta_0$=10.56° @3.55 GHz.

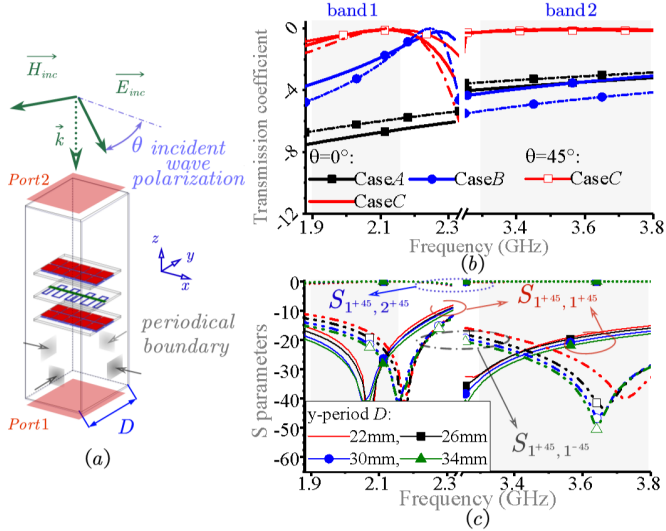

**Fig. 2.** (a) Configuration of the simulation model, (b) Transmission coefficient for *Case*-A to *Case*-C (solid line: full wave model, dash-dot line: equivalent circuit model), (c) *S* parameters for *Case*-C with varies y-period *D*.

To validate the equivalent circuit and further reveal the operating mechanism, the transmission coefficients are compared among *Case*-A to *Case*-C. The simulation model, as shown in Fig. 2(a), uses a periodic boundary surrounding the structures. The angle between the polarization direction of the incident field and the *x*-axis is $\theta$. Fig. 2(b) shows that when $\theta$=0°, *Case*-A exhibits poor performance with a transmission loss exceeding 3 dB, while *Case*-B has a single passband near 2.3 GHz; *Case*-C has two distinct passbands denoted as *band* 1 and *band* 2, spanning 1.92-2.18 GHz and 3.3-3.8 GHz, respectively, with a transmission coefficient over -0.92 dB. Meanwhile, the transmission coefficients of the equivalent circuit (dash-dot lines) closely match those of the full-wave simulation (solid lines), verifying the circuit's accuracy. Therefore, the *band* 1 is created by the interaction between the thin strip and the periodic meander line, while *band* 2 is generated by the capacitive patch. These results suggest that additional passbands can be introduced by incorporating more serial *L-C* tanks.

Notably, although the proposed DBTS resonates only along the *x*-axis, it also exhibits a high transmission coefficient for an incident wave polarized in $\theta$=45°, as shown in Fig. 2(b). This is attributed to the small electromagnetic size along the non-resonant *y*-axis. Meanwhile, Fig. 2(c) compares the *S* parameters of the DBTS for different periodic boundary widths *D* in the y direction. The resonance frequencies remain nearly unchanged with varying y-periods, indicating that increasing the y-period has minimal impact on the performance of the DBTS.

### III. TRI-BAND APERTURE-SHARED ANTENNA ARRAY

To address the shielding effect caused by the LB antenna in TBAs, the proposed DBTS is used to construct an LB dipole antenna, which is then integrated with a dual-band sub-array operating at MB and HB to form a TBA.

#### A. Configuration of the TBA

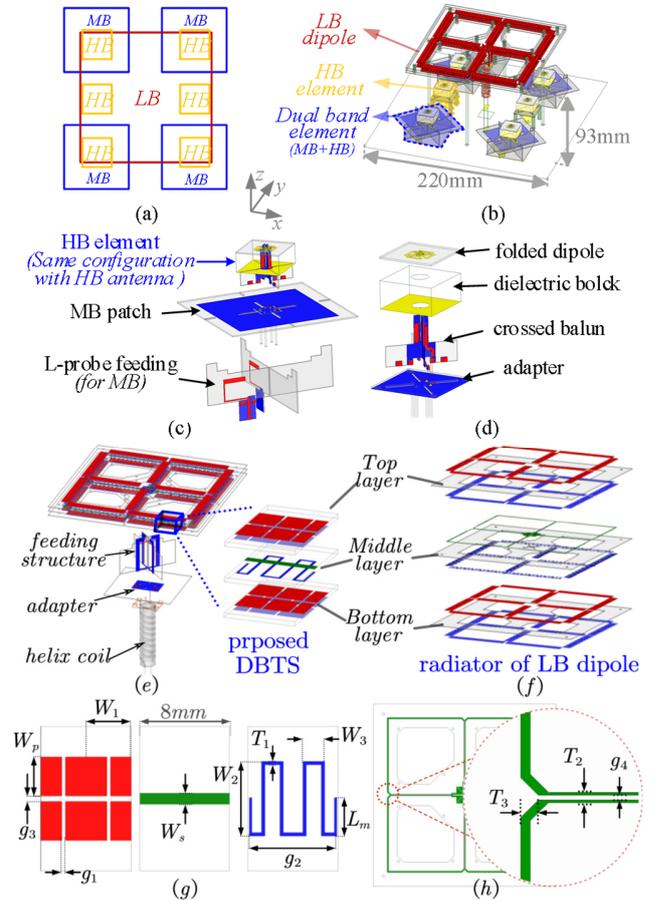

**Fig. 3.** Configuration of the TBA: (a) Topology sketch, (b) 3D view. Configurations for: (c) The dual-band antenna, (d) The high-band antenna. (e) The LB antenna (f) LB radiator. The dimensions of: (g) The DBTS and (h) Top layer of the LB radiator. The key parameters: $W_1$=2.95mm, $W_2$=6.7mm, $W_3$=1.93mm, $W_5$=20mm, $W_p$=3.5mm, $W_s$=1mm, $L_m$=3.45mm, $r_1$=5mm, $r_2$=3mm, $g_1$=0.25mm, $g_2$=7.75mm, $g_3$=$g_4$=0.5mm, $T_1$=0.3mm, $T_2$=1.1mm, $T_3$=2mm.

As shown in Fig. 3(a-b), the tri-band antenna array consists of an LB dipole and a dual-band sub-array operating in the MB and HB. The dual-band sub-array is arranged with four dual-band antennas interleaved with two HB antennas. The array spacing is 100 mm in the x-direction. In the y-direction, the spacing is 60 mm for the HB while 120 mm for the MB, respectively. The configurations of the dual-band and the HB antenna are shown in Fig. 3(c-d). The dual-band antenna consists of one HB element stacked upon one MB patch antenna, while the HB element shares the same configuration with the HB antenna shown in Fig. 3(d). The radiator of LB dipole is illustrated in Fig. 3(e-h), which is constructed by the proposed DBTS to mitigate the shielding effect. Noting that the DBTS with coupled line is employed in the middle part of the dipole, the coupled-line width is appropriately designed so that the coupled-line DBTS operates within the same passbands as the single-line one. All the substrates used in the antenna array design are Rogers Kappa438 laminates with thickness of 0.508mm and permittivity of 4.38.

*B. Scattering Suppression Effect of the Proposed LB Dipole*

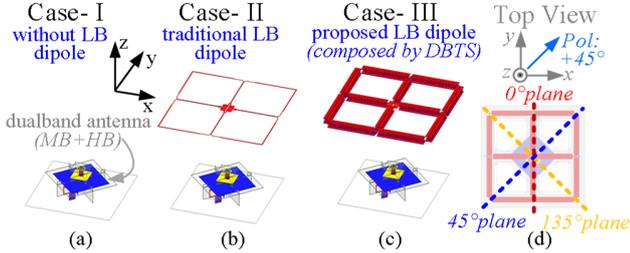

**Fig. 4.** Configurations of the three cases: (a) *Case*-I: without LB dipole, (b) *Case*-II: traditional LB dipole, (c) *Case*-III: LB dipole composed by the DBTS; (d) Schematic on observation plane cuts in top view.

To demonstrate the scattering suppression effect of the proposed LB antenna at MB and HB, three cases are considered, as illustrated in Fig. 4. In *Case*-I, only the dual-band antenna operating at MB and HB is considered. In *Case*-II, a traditional LB dipole is set above the dual-band antenna coaxially. The *Case*-III is like *Case*-II, but using the LB dipole composed of the DBTS instead. The 3D radiation patterns for the three cases are compared in Fig. 5. Compared to *Case*-I, radiation pattern deterioration is observed in *Case*-II, indicating a significant shielding effect caused by the traditional LB dipole. In contrast, the 3D radiation patterns in *Case*-I and *Case*-III are highly similar for both MB and HB. This similarity demonstrates that the proposed LB dipole effectively suppresses the shielding effect, preserving the radiation pattern as if no LB dipole were present.

To quantitatively evaluate the reduction in shielding effect, Fig. 6 compares the radiation patterns for these three cases. Considering the feature of symmetry, only three plane cuts (azimuth angle = 0°, 45°, and 135°) of the patterns are illustrated. In *Case*-II, the broadside gain is about 2 dB lower than in *Case*-I for MB, and approximately 1 dB lower than in *Case*-I for HB. In contrast, the radiation patterns of *Case*-III are very similar to those of *Case*-I, with a broadside gain difference of less than 0.5 dB for both MB and HB. Therefore, incorporating the DBTS with the dipole effectively restores the radiating performance.

Noting that although the DBTS is designed under periodic boundary conditions, and the DBTS in the LB dipole operates with an open boundary in the non-resonant direction, the passband characteristics remain nearly unchanged in both cases. This is because the DBTS performance is not sensitive to the non-resonant direction period, while the open boundary represents the limiting case of an infinitely large period. Consequently, the DBTS designed under periodic boundaries remains suitable for integration into the LB dipole [18].

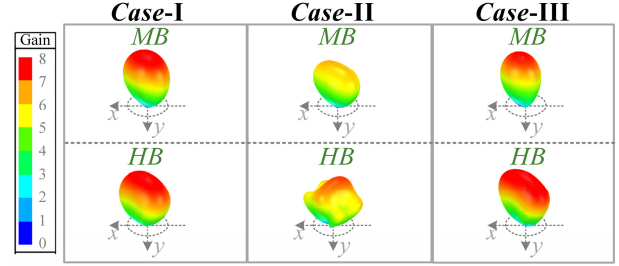

**Fig. 5.** 3D radiation patterns at HB (3.55 GHz) and MB (2.05 GHz) for *Case*-I to *Case*-III.

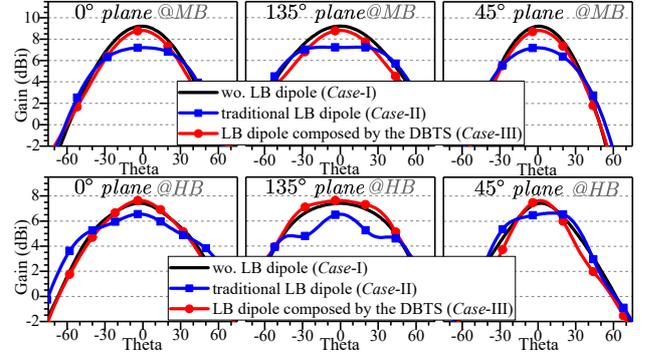

**Fig. 6.** Radiation patterns for *Case*-I, II, III when the azimuth angle is respectively equal to 0°, 45° and 135° at HB and MB.

## IV. RESULTS AND DISCUSSION

To validate the design, a prototype of the proposed TBA was fabricated and measured. The prototype and ports distribution are shown in Fig. 7(a). The S Parameters of the proposed antenna array is measured with R&S-ZVA40 vector network analyzer. Only the performance of ports $H_1^{+-}$, $H_2^{+-}$, $M_1^{+-}$, $L^{+-}$ are shown due to array symmetry. As shown in Fig. 7(b), the measured reflection coefficients of ports $L^{+-}$, $M_1^{+-}$, $H_1^{+-}$ and $H_2^{+-}$ are lower than -10 dB, showing good agreement with the simulation results. Radiation patterns were measured in an MVG near-field chamber. Since the array is intended for MIMO applications, column-wise feeding was applied during measurement. The measured broadside gains for each operating band are shown in the right side of Fig. 7(b), with discrepancies of 0.5 to 1 dB compared to simulations, primarily due to cable and adaptor losses. Measured renormalized radiation patterns at the center frequencies of the three bands are shown in Fig. 8. Slight asymmetry is observed in the H-plane radiation pattern at HB, attributed to the asymmetric placement of HB antennas on the ground plane. However, this asymmetry has negligible impact on the broadside gain and 3-dB beamwidth, thus not affecting the practical applicability in base-station deployments.

TABLE I
COMPARISON BETWEEN CURRENTLY MULTIBAND APERTURE-SHARED ANTENNA ARRAYS

| Works | Frequency bands (GHz) | Shielding effect reduction method | Max. broadside gain deviation * (dB) | Num. of elements | Aperture reuse ratio | Aperture-shared area ($\lambda_L^2$) |
|---|---|---|---|---|---|---|
| [4] | 0.69–0.96 (32.7%) 1.8–2.7 (40%) 3.3–3.8 (14%) | high-stop surface + top-bottom scheme | Not available | HB: 16, MB: 4 | HB: 54.1%, MB: 69.9%" | 0.4356 |
| [8] | 0.79-0.96 (19.4%) 1.71-2.17 (23.7%) 3.4-3.6 (5.7%) | SSR + choke | ~0.55 (MB) ~1.23 (HB) | HB: 24 MB: 4 | HB: 27.7%, MB: 4% | 0.466 |
| [22] | 0.69-0.96 (32.7%) 1.7-2.4 (34.1%) | slot loading | ~0.83 | 4 | 56.70% | 0.259 |
| [23] | 0.68-1.07 (44.57%) 1.7-2.7 (45.45%) | quasi F-P cavity | ~1.04 | 4 | 41.25% | 0.286 |
| [7] | 0.698-0.96 (31.6%) 1.7-2.17 (24.29%) | choke | ~0.7 | 4 | 27.7% | 0.304 |
| [15] | 1.8-2.7 (40%) 3.3-3.8 (14%) | Slot-ring FSS | ~0.64 | 4 | 17.67% | 0.345 |
| This work | 0.65-0.88 (30%) 1.92-2.18 (12.7%) 3.3-3.8 (14%) | FSS-based DBTS | 0.57 (MB) 0.5 (HB) | HB: 6 MB: 4 | HB: 100% MB: 51% | 0.203 |

*The broadside gain deviation of the other works is evaluated from the figure in their articles.

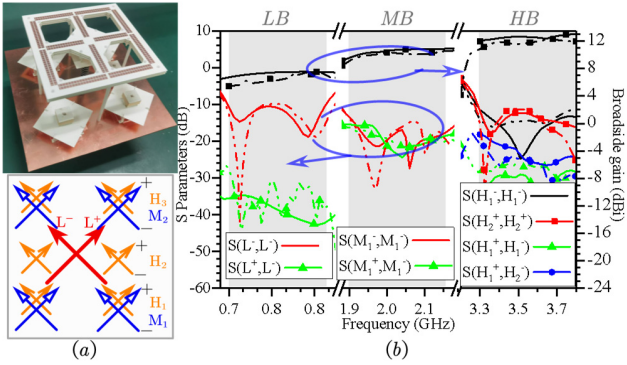

**Fig. 7.** (a) Porotype and port distribution of the proposed array; (b) Reflection coefficient and isolation within same element for ports $L^{+-}$, $M_1^{+-}$, $H_1^{+-}$ and $H_2^{+-}$ (left side), along with broadside gains (right side) of the array for each band. (Note that simulated results are presented with solid lines while measured results are in dashed-dot lines)

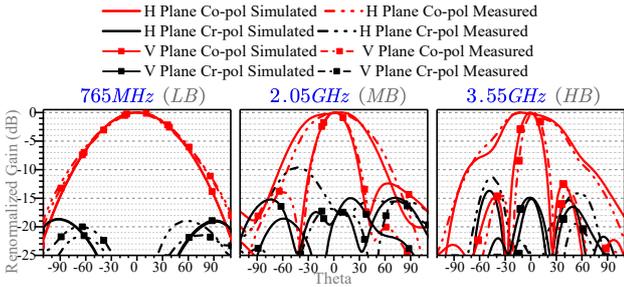

**Fig. 8.** Simulated and measured renormalized radiation pattern in horizontal and vertical plane at LB, MB and HB, respectively.

For further demonstration, a comparison with other reported multiband aperture-shared antenna arrays is shown in Table. I. The aperture reuse ratio quantifies the extent of aperture overlap among antennas operating at different frequency bands, and is defined as

$$aperture\ reuse\ ratio = \frac{A_{shared}}{A_{band}} \quad (1)$$

where $A_{shared}$ denotes the total area shared among antennas with overlapping apertures, and $A_{band}$ represents the total aperture area of the corresponding frequency band. The aperture-shared area, on the other hand, refers to the physical area where the apertures of different-band antennas spatially overlap. A smaller shared area generally implies a more compact structure and higher integration level. In this work, the proposed array achieves a 100% aperture reuse ratio at HB and 51% at MB, along with the smallest aperture-shared area. The array also exhibits minimal broadside gain deviation. While the array in [4] offers a higher MB reuse ratio and broader bandwidth, its use of helix-coil feeding lines for HB and MB introduces additional insertion loss. Also, mounting the HB and MB antennas atop the LB element leads to structural instability. TBA in [8] achieves wider MB bandwidth, but exhibits narrower bandwidths at HB and LB. Its use of choke loading suppresses LB scattering but limits LB bandwidth, while the MB antennas are placed far from the LB element, leading to a lower reuse ratio and increased aperture-shared area. The dual-band array in [22] attains a reuse ratio of 56.7%, but with a gain deviation of 0.83 dB. The design in [23] covers a wide bandwidth but suffers from a gain deviation of 1.04 dB. Arrays in [7] and [15] achieve small gain deviations, yet their aperture-shared areas are considerably larger. Overall, the proposed design offers a feasible solution to the multiband base station application, achieving high integration, stable structure, while maintaining well-preserved radiation performance at higher bands.

V. CONCLUSION

To alleviate the shielding effect of the LB dipole in a TBA, a DBTS with frequency bands scalability is proposed in this work, and is used to construct the LB element. By integrating it with a dual-band MB/HB sub-array, a compact TBA is realized. As a result, the radiation performance of the MB and HB antennas is effectively restored, with the broadside gain deviation caused by the shielding effect below 0.6 dB.

References

[1] Y. Zhu, Y. Chen, and S. Yang, "Helical Torsion Coaxial Cable for Dual-Band Shared-Aperture Antenna Array Decoupling," *IEEE Trans. Antennas Propag.*, vol. 68, no. 8, pp. 6128–6135, Aug. 2020, doi: 10.1109/TAP.2020.2986725.

[2] Y. Chen, J. Zhao, and S. Yang, "A Novel Stacked Antenna Configuration and its Applications in Dual-Band Shared-Aperture Base Station Antenna Array Designs," *IEEE Trans. Antennas Propag.*, vol. 67, no. 12, pp. 7234–7241, Dec. 2019, doi: 10.1109/TAP.2019.2930136.

[3] Y. Zhu, Y. Chen, and S. Yang, "Decoupling and Low-Profile Design of Dual-Band Dual-Polarized Base Station Antennas Using Frequency-Selective Surface," *IEEE Trans. Antennas Propag.*, vol. 67, no. 8, pp. 5272–5281, Aug. 2019, doi: 10.1109/TAP.2019.2916730.

[4] D. He, Y. Chen, and S. Yang, "A Low-Profile Triple-Band Shared-Aperture Antenna Array for 5G Base Station Applications," *IEEE Trans. Antennas Propag.*, vol. 70, no. 4, pp. 2732–2739, Apr. 2022, doi: 10.1109/TAP.2021.3137486.

[5] X. W. Dai, C. Ding, F. Zhu, L. Liu, and G. Q. Luo, "Broadband Dual-Polarized Element With Rotated Sleeve Arms for Compact Dual-Band Antenna," *IEEE Antennas Wirel. Propag. Lett.*, vol. 20, no. 12, pp. 2519–2523, Dec. 2021, doi: 10.1109/LAWP.2021.3117037.

[6] H.-H. Sun, H. Zhu, C. Ding, B. Jones, and Y. J. Guo, "Scattering Suppression in a 4G and 5G Base Station Antenna Array Using Spiral Chokes," *IEEE Antennas Wirel. Propag. Lett.*, vol. 19, no. 10, pp. 1818–1822, Oct. 2020, doi: 10.1109/LAWP.2020.3019930.

[7] H.-H. Sun, C. Ding, H. Zhu, B. Jones, and Y. J. Guo, "Suppression of Cross-Band Scattering in Multiband Antenna Arrays," *IEEE Trans. Antennas Propag.*, vol. 67, no. 4, pp. 2379–2389, Apr. 2019, doi: 10.1109/TAP.2019.2891707.

[8] Y.-L. Chang and Q.-X. Chu, "Suppression of Cross-Band Coupling Interference in Tri-Band Shared-Aperture Base Station Antenna," *IEEE Trans. Antennas Propag.*, vol. 70, no. 6, pp. 4200–4214, Jun. 2022, doi: 10.1109/TAP.2021.3138531.

[9] W. Niu, B. Sun, G. Zhou, and Z. Lan, "Dual-Band Aperture Shared Antenna Array With Decreased Radiation Pattern Distortion," *IEEE Trans. Antennas Propag.*, vol. 70, no. 7, pp. 6048–6053, Jul. 2022, doi: 10.1109/TAP.2022.3161267.

[10] H.-H. Sun, B. Jones, Y. J. Guo, and Y. H. Lee, "Suppression of Cross-Band Scattering in Interleaved Dual-Band Cellular Base-Station Antenna Arrays," *IEEE Access*, vol. 8, pp. 222486–222495, 2020, doi: 10.1109/ACCESS.2020.3043578.

[11] S. J. Yang, R. Ma, and X. Y. Zhang, "Self-Decoupled Dual-Band Dual-Polarized Aperture-Shared Antenna Array," *IEEE Trans. Antennas Propag.*, vol. 70, no. 6, pp. 4890–4895, Jun. 2022, doi: 10.1109/TAP.2021.3137531.

[12] W. Niu, B. Sun, and X. Huang, "A Filtering and Electromagnetic-Transparent Antenna for Triple-Band Aperture-Shared Base Station Antenna Array," *IEEE Antennas Wirel. Propag. Lett.*, vol. 23, no. 1, pp. 244–248, Jan. 2024, doi: 10.1109/LAWP.2023.3322414.

[13] Y. Da, X. Chen, and A. A. Kishk, "In-Band Mutual Coupling Suppression in Dual-Band Shared-Aperture Base Station Arrays Using Dielectric Block Loading," *IEEE Trans. Antennas Propag.*, vol. 70, no. 10, pp. 9270–9281, Oct. 2022, doi: 10.1109/TAP.2022.3177496.

[14] P. Liu, F. Jia, Y. Zhang, G. Su, Q. Wang, and X. Y. Zhang, "Dual-Polarized Dipole Antenna With Dual-Band Spatial Filtering Response for Aperture-Shared Triband Base Station Array Application," *IEEE Antennas Wirel. Propag. Lett.*, vol. 22, no. 12, pp. 3057–3061, Dec. 2023, doi: 10.1109/LAWP.2023.3309876.

[15] D. He, Q. Yu, Y. Chen, and S. Yang, "Dual-Band Shared-Aperture Base Station Antenna Array With Electromagnetic Transparent Antenna Elements," *IEEE Trans. Antennas Propag.*, vol. 69, no. 9, pp. 5596–5606, Sep. 2021, doi: 10.1109/TAP.2021.3061151.

[16] Y. Zhu, Y. Chen, and S. Yang, "Cross-Band Mutual Coupling Reduction in Dual-Band Base-Station Antennas With a Novel Grid Frequency Selective Surface," *IEEE Trans. Antennas Propag.*, vol. 69, no. 12, pp. 8991–8996, Dec. 2021, doi: 10.1109/TAP.2021.3098514.

[17] G.-N. Zhou, B.-H. Sun, Q.-Y. Liang, S.-T. Wu, Y.-H. Yang, and Y.-M. Cai, "Triband Dual-Polarized Shared-Aperture Antenna for 2G/3G/4G/5G Base Station Applications," *IEEE Trans. Antennas Propag.*, vol. 69, no. 1, pp. 97–108, Jan. 2021, doi: 10.1109/TAP.2020.3016406.

[18] S. J. Yang, Y. Yang, and X. Y. Zhang, "Low Scattering Element-Based Aperture-Shared Array for Multiband Base Stations," *IEEE Trans. Antennas Propag.*, vol. 69, no. 12, pp. 8315–8324, Dec. 2021, doi: 10.1109/TAP.2021.3083760.

[19] X. Liu *et al.*, "A Mutual-Coupling-Suppressed Dual-Band Dual-Polarized Base Station Antenna Using Multiple Folded-Dipole Antenna," *IEEE Trans. Antennas Propag.*, vol. 70, no. 12, pp. 11582–11594, Dec. 2022, doi: 10.1109/TAP.2022.3209177.

[20] Y. Jia, H. Zhai, J. Yin, Y. Wang, and Y. Liu, "A Quadruple-Band Shared-Aperture Antenna Array With Multiband Radiation Pattern Restorations," *IEEE Trans. Antennas Propag.*, vol. 72, no. 10, pp. 7722–7735, Oct. 2024, doi: 10.1109/TAP.2024.3439874.

[21] Y. Wang, H. Su, R. C. Dai, L. T. Chen, L. H. Ye, and X. Y. Zhang, "A Systematic Approach to Design Spatial Filtenna for Aperture-Shared Base-Station Array," *IEEE Trans. Antennas Propag.*, vol. 71, no. 10, pp. 7792–7803, Oct. 2023, doi: 10.1109/TAP.2023.3299409.

[22] Y.-S. Wu, Q.-X. Chu, and H.-Y. Huang, "Electromagnetic Transparent Antenna With Slot-Loaded Patch Dipoles in Dual-Band Array," *IEEE Trans. Antennas Propag.*, vol. 70, no. 9, pp. 7989–7998, Sep. 2022, doi: 10.1109/TAP.2022.3164194.

[23] Y. Qin, R. Li, Q. Xue, X. Zhang, and Y. Cui, "Aperture-Shared Dual-Band Antennas With Partially Reflecting Surfaces for Base-Station Applications," *IEEE Trans. Antennas Propag.*, vol. 70, no. 5, pp. 3195–3207, May 2022, doi: 10.1109/TAP.2021.3137448.

[24] S.-Y. Sun, C. Ding, W. Jiang, and Y. J. Guo, "Simultaneous Suppression of Cross-Band Scattering and Coupling Between Closely Spaced Dual-Band Dual-Polarized Antennas," *IEEE Trans. Antennas Propag.*, vol. 71, no. 8, pp. 6423–6434, Aug. 2023, doi: 10.1109/TAP.2023.3284136.

[25] R. C. Dai, H. Su, S. J. Yang, J.-H. Ou, and X. Y. Zhang, "Broadband Electromagnetic-Transparent Antenna and Its Application to Aperture-Shared Dual-Band Base Station Array," *IEEE Trans. Antennas Propag.*, vol. 71, no. 1, pp. 180–189, Jan. 2023, doi: 10.1109/TAP.2022.3217351.

[26] S. M. A. Momeni Hasan Abadi, M. Li, and N. Behdad, "Harmonic-Suppressed Miniaturized-Element Frequency Selective Surfaces With Higher Order Bandpass Responses," *IEEE Trans. Antennas Propag.*, vol. 62, no. 5, pp. 2562–2571, May 2014, doi: 10.1109/TAP.2014.2303822.

[27] F. Costa, A. Monorchio, and G. Manara, "An Overview of Equivalent Circuit Modeling Techniques of Frequency Selective Surfaces and Metasurfaces," vol. 29, no. 12, 2014.